\documentclass{aa}

\usepackage{graphicx}

\def\arcdeg{\hbox{$^\circ$}}

\begin{document}

\thesaurus{11.03.1;11.04.1;11.05.02;11.06.2;11.09.2} 
 
\title{On the $L_x - \sigma_v$ relation of groups of galaxies}

\author{Manolis Plionis
\inst{1,2}
\and
Hrant M. Tovmassian
\inst{1}
}

\offprints{M.Plionis}

\institute{Instituto Nacional de Astrof\'{i}sica \'Optica y
Electr\'onica, AP 51 y 216, 72000, Puebla, Pue, M\'exico
email: mplionis@inaoep.mx
\and
Institute of Astronomy \& Astrophysics, National Observatory of
Athens, I.Metaxa \& B.Pavlou, P.Penteli 152 36, Athens, Greece
}

\titlerunning{X-ray emission of galaxy groups} 
\authorrunning{M. Plionis \& H.M. Tovmassian}

\date{Received ...2003 / Accepted  .. 2003}

\maketitle

\abstract{ 
We analyse the $L_x-\sigma_v$ relation for the new Mulchaey et al.
group Atlas. We find that once we take into account the possible
statistical bias introduced by the cutoff in luminosity, we recover a
relation which is consistent with that of clusters, ie., 
$L_x \propto \sigma^{4.1}$. The larger scatter of this relation for
groups of galaxies could be attributed to an orientation effect, 
due to which the radial velocity 
dispersion of groups oriented close to orthogonal to the line of
sight, would be underestimated. This effect could also contribute in
the direction of flattening the slope of the group
$L_x-\sigma_v$ relation.

}

\maketitle

\section{Introduction}

Most galaxies in the universe occur in small groups (cf. Geller \& Huchra 1983; 
Tully 1987; Nolthenius \& White 1987), which in many respects could be 
considered as poor clusters of galaxies. The detection of X-ray emission from 
some groups has increased considerably the interest to study them
since they proved to be real entities and not projection effects (for
an extensive review see Mulchaey 2000).
Solinger \& Tucker (1972) showed that if the source of the X-ray emission is 
hot gas bound in clusters, then the X-ray luminosity, $L_{x}$, should be 
correlated with the optical radial velocity dispersion, $\sigma_v$. 
Simple theoretical arguments show that in a virialized, isothermal
aggregation of gas, which emits
thermal bremsstrahlung emission ($L_{x} \propto \int \rho_{gas}^2
T(r)^{1/2} {\rm d}V$), we can obtain that $L_x \propto M^{4/3}$ (where
$M$ is the total of the system) and using
virial arguments ($M \propto \sigma^3$) 
we then have that $L_x$ should be roughly 
proportional to the fourth power of $\sigma$: $\log L_{x} \propto 
\log \sigma_v^{4}$ (cf. Navaro, Frenk \& White 1995). 
Quintana \& Melnick (1982) first showed that the X-ray 
luminosity of clusters of galaxies obey the expected correlation. 

Numerical simulations have shown that the
relationship between $L_x$ and $\sigma_v$ for groups should be similar to 
that of clusters (cf. Navarro, Frenk, \& White 1997), or even steeper
if one takes into account radiative cooling which significantly
reduces the amount of the hot gas fraction at low-$\sigma_v$'s (Dav\'e, Katz
\& Weinberg 2002). Such a steep relation for systems with
$L_x<10^{43}$ ergs s$^{-1}$, has been advocated by Mahdavi \& Geller
(2001), although they notice an erratic behavior of the poor groups
of galaxies.

In some observational studies a consistency has been found between the
$L_x-\sigma_v$ relation of groups and clusters with a slope $\sim 4$
(cf. Ponman et al 1996; Mulchaey \& Zabludoff 1998; Helsdon \& Ponman
2000), while in others a shallower slope has been found, 
i.e., groups appear to have a relatively enhanced X-ray emission to what 
predicted from the $L_x - \sigma_v$ relation deduced from clusters of 
galaxies (cf. Dell`Antonio et al. 1994; Mahdavi et al. 1997, 2000; 
Xue \& Wu 2000). 
Attempts to explain such deviations of the group from the cluster
$L_x-\sigma_v$ behaviour have invoked a possible contribution of 
individual galaxy halos to the group X-ray luminosity (Mahdavi et
al. 2000), poorly determined $\sigma_v$'s and/or $L_x$'s (Zimer,
Mulchaey \& Zabludoff 2001), a large scatter due to
an non-equilibrium galaxy velocity distribution (Mahdavi \& Geller 2001).

An alternative explanation, 
based on a possible orientation effect, was proposed by Tovmassian, Tiersch, 
\& Yam (2002) [see also Tovmassian, Martinez, \& Tiersch 1999; Tovmassian 2002]. 
They considered the relatively small RASSCALS and HCG 
X-ray group samples (Mahdavi et al. 2000; Ponman et al. 1996, respectively)
and showed that the flattening of groups with relatively 
small velocity dispersions is, on average, larger than those of groups with 
higher velocity dispersions. They suggested that the shallow shape of $\log 
L_{x}-\log \sigma_v$ of groups of galaxies could be partly the
result of an underestimation of the velocity dispersion
of elongated groups when seen roughly orthogonally to the line
of sight, while when seen edge-on, they will have
higher and probably nearer to their true $\sigma_v$ values.
This correlation could be explained if member galaxies in 
groups move preferentially along the group elongation,
since groups have been found to have a prolate-like shape 
(Hickson et al. 1984; Malykh \& Orlov 1986; Oleak 
et al. 1998) as in the case of clusters (cf. Plionis, 
Barrow \& Frenk 1991).

Recently Mulchaey et al. (2003) published an X-ray Atlas of groups of 
galaxies, which is the largest sample of groups studied to date having
X-ray ROSAT PSPC pointed observations. In this paper we
address two questions: is there sufficient evidence to support claims that
groups have enhanced, with respect to clusters, X-ray emission (ie.,
that the power law index of the $\log L_x - \log\sigma^\alpha$ relation
has $\alpha <<4$) and
what is the origin of the larger scatter of the $L_x - \sigma$ relation
for groups.

\section {Data and results}
In order to address the issue of the possibly enhanced group X-ray emission
we have selected from the Atlas of Mulchaey et al (2003) 
those with detected X-ray emission
that contain less than 20 members (since a larger membership should
rather define poor clusters). Note that the $L_{x}$ values were
determined after the removal of point sources.
We exclude three groups: NGC 2484 and NGC 6251 because they consist of 
only two members and the NGC 
6329 group due to its very large projected length $a$ 
($\approx~11$Mpc)\footnote{Lengths are determined assuming $H_{o}$ = 72 km 
s$^{-1}$Mpc$^{-1}$.}, which is suspected that it could be a 
superposition of two or even more groups and hence, its $\sigma_v$ 
value may be unreliable. We are left with 43 groups in total out of
which we have three triplets. Although, Focardi \& Kelm (2002) argue
that triplets are a distinct class of low-$\sigma_v$ objects, in
our case they span the whole $\sigma_v$ range.

In Fig. 1 we plot $\log L_{x}$ versus $\log \sigma_v$ for the considered 
groups (open symbols) and for groups with members $N>20$ (squares), 
which could be considered as poor clusters. The latter are
closely located along the line $L_{x} \propto \sigma_v^{4}$ (solid line). 

\noindent
We perform a direct least-square regression to the $N<20$ groups of the type 
\begin{equation}\label{eq:1}
\log L_x = \alpha \log\sigma + C_1
\end{equation} 
and find a strong correlation with a correlation coefficient $R=0.65$
and a probability of random occurrence $3\times 10^{-6}$. The fitted
parameters are:
$$ \alpha=1.72 \pm 0.31 \;\;\;\;\;\;\;\ C_1 = 37.6 \pm 0.75 \;.$$ 
The determined slope is much shallower than the expected value for clusters
$\alpha\simeq 4$, while the value found from the $N>20$
groups is $\alpha \simeq 3.2$.
As discussed in the introduction, the shallower slope of the 
group $L_x-\sigma_v$ relation already been noticed in other studies as
well and indeed it has stimulated speculations on the 
reason for the apparently enhanced X-ray emission of groups. 
Hence, we could have concluded from our results that
this sample supports the claims for a relatively enhanced X-ray emission for
groups. 
\begin{figure}[t]
\includegraphics[width=9cm]{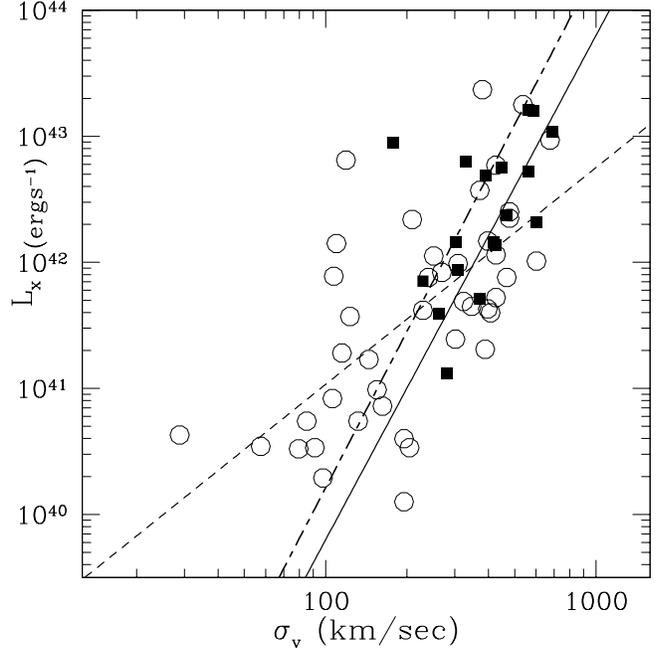}
\caption{$L_x - \sigma_v$ relation for groups: those with $N\le20$ are 
represented as circles, while those with $N>20$, supposed to be poor 
clusters, are marked by filled squares. The solid line is the $\log L_{x} 
\propto \log \sigma_v^{4}$ relation of clusters of galaxies. The dashed line
is the best direct regression fit for the groups 
($\log L_{x} \propto \log \sigma_v^{1.7}$) and the dot-dashed line is
is the best inverse regression fit for the groups 
($\log \sigma \propto (1/4.1) \log L_{x}$).} 
\end{figure}

\subsection{Statistical bias ?}
However, what we are witnessing is the result of a statistical bias,
resembling the Malmquist bias, which appears because of the low-$L_x$ limit (either
due to the lower mass limit necessary for the ICM to light up, or due
to the X-ray flux limit in the construction of the sample). This bias 
is of the same nature that enters in scaling relation
(eg. Tully-Fisher, Faber-Jackson, etc) 
where the magnitude or flux limit imposes a bias
such that the slope of the derived relation is shallower than the
nominal one. The larger the scatter of the relation, the
larger the bias imposed. This bias may or may not appear in the corresponding
cluster relation depending on the amplitude of the scatter around the nominal
relation. To understand this bias let us imagine that the cluster
relation $L_x \propto \sigma^4$ is obeyed by groups as well
and let us consider a subset of our sample of groups that have
$\sigma_v \simeq 600$ km s$^{-1}$ 
so that they are typically quite
brighter than the lower $L_x$ limit. It is clear that their $\log L_x$
value will be distributed around the mean $\langle \log L_x \rangle$ 
value with some dispersion $\sigma$. However, if we take a
subsample with $\sigma_v \simeq 100$ km s$^{-1}$ so that they are
on average quite
faint with $\langle \log L_x \rangle \simeq \log (L_x)_{limit}$, then
the only groups that
will appear in the sample are those with $\log L_x > \langle \log L_x
\rangle$ and none with $\log L_x < \langle \log L_x \rangle$. This
will induce the above mentioned bias.

To see this more clearly, we have performed Monte-Carlo simulations in
which we assume a relation $L_x \propto \sigma_v^{4}$, and a
Gaussian scatter of $\delta \log L_x = 0.8$.
We then perform a direct (ie., eq.\ref{eq:1}) 
and an inverse least-square regression fit to the
resulting data (ie., we fit $\log\sigma_v = (1/\alpha) \log L_x +
C_2$). In Fig. 2a we show such a simulation with 5000 ``groups'',
where the solid line is the input $L_x \propto \sigma^4$ relation and
the dashed line is the recovered from the direct regression. The inverse
regression recovers exactly the input slope.

Furthermore, to study more accurately the effect of this statistical
bias on our sample we have performed a series of simulations in which 
we use the observed $L_x$ values of the groups and we derive the values
$\sigma_v$, assuming a Gaussian scatter of the $\log L_x - \log\sigma_v$
relation of varying magnitude. 
We have performed 1000 such simulations for each different value of the
Gaussian scatter. In Fig. 2b we plot the derived values of the slope
$\alpha$ for both regression methods as a function input scatter. 
The vertical line indicates the scatter of the relation derived from the $N\le 20$
sample of groups.
The direct regression method (star-like symbols) underestimates
severely the input slope, with the underestimation increasing with
increasing scatter, while the inverse regression method (open
symbols) recovers it accurately. 
Note also that we
plot (squares) the results of the inverse regression in case that
there is a sharp cutoff at $\sigma_v=1000$ km s$^{-1}$ , in which case a
similar, although of smaller magnitude, bias is introduced in the
opposite direction.

It is evident that:
\begin{itemize}
\item The inverse regression recovers correctly the input slope of the
$\log L_x - \log \sigma_v$ relation for all values of the scatter.
\item for the observed amount of scatter ($\delta C_1 \simeq 0.8$) the
expected slope of the relation once we use the direct regression is
around $\sim 2$, as indeed observed.
\end{itemize}

\begin{figure*}[t]
\includegraphics[width=9cm]{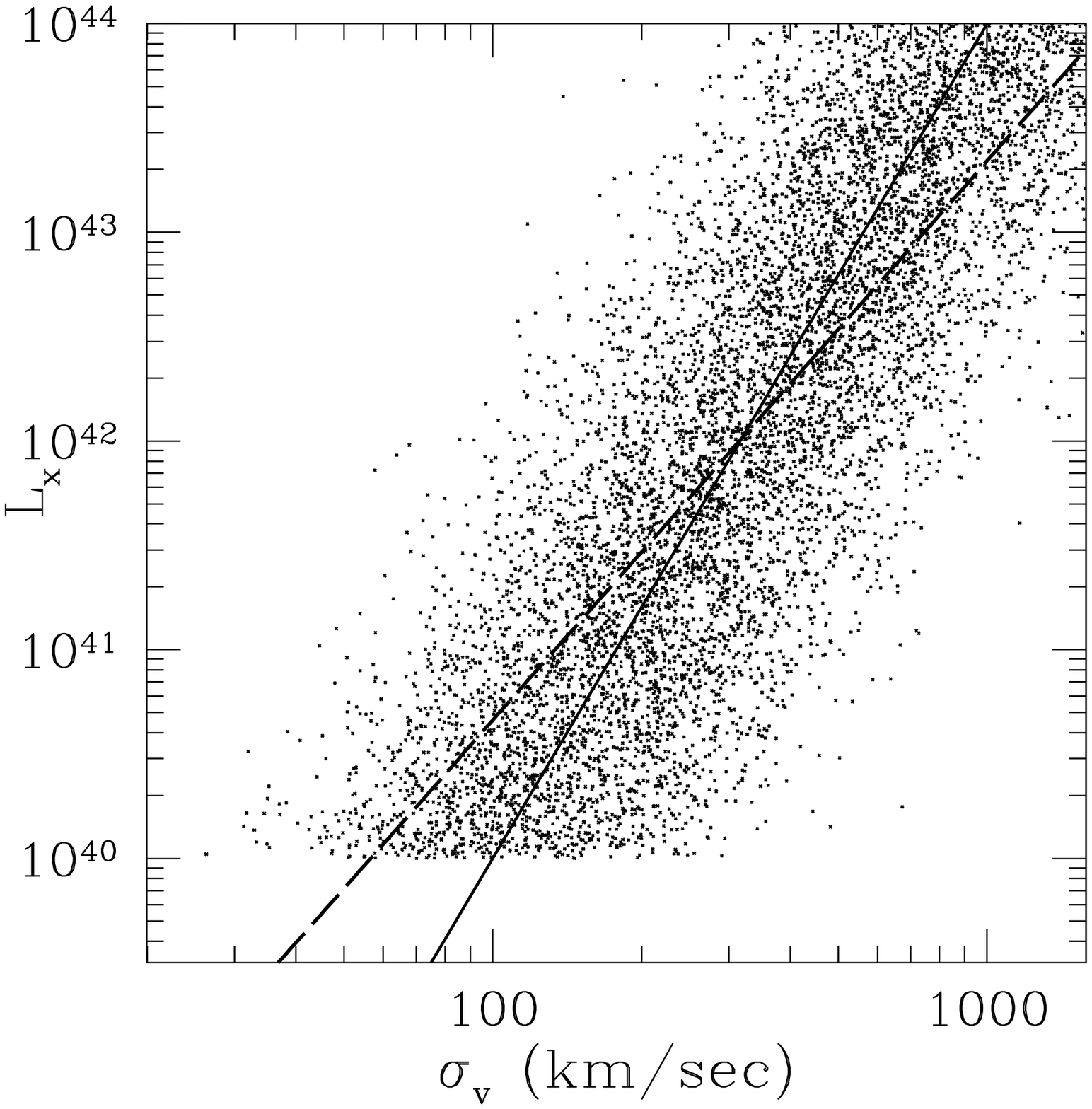} \hfill
\includegraphics[width=9cm]{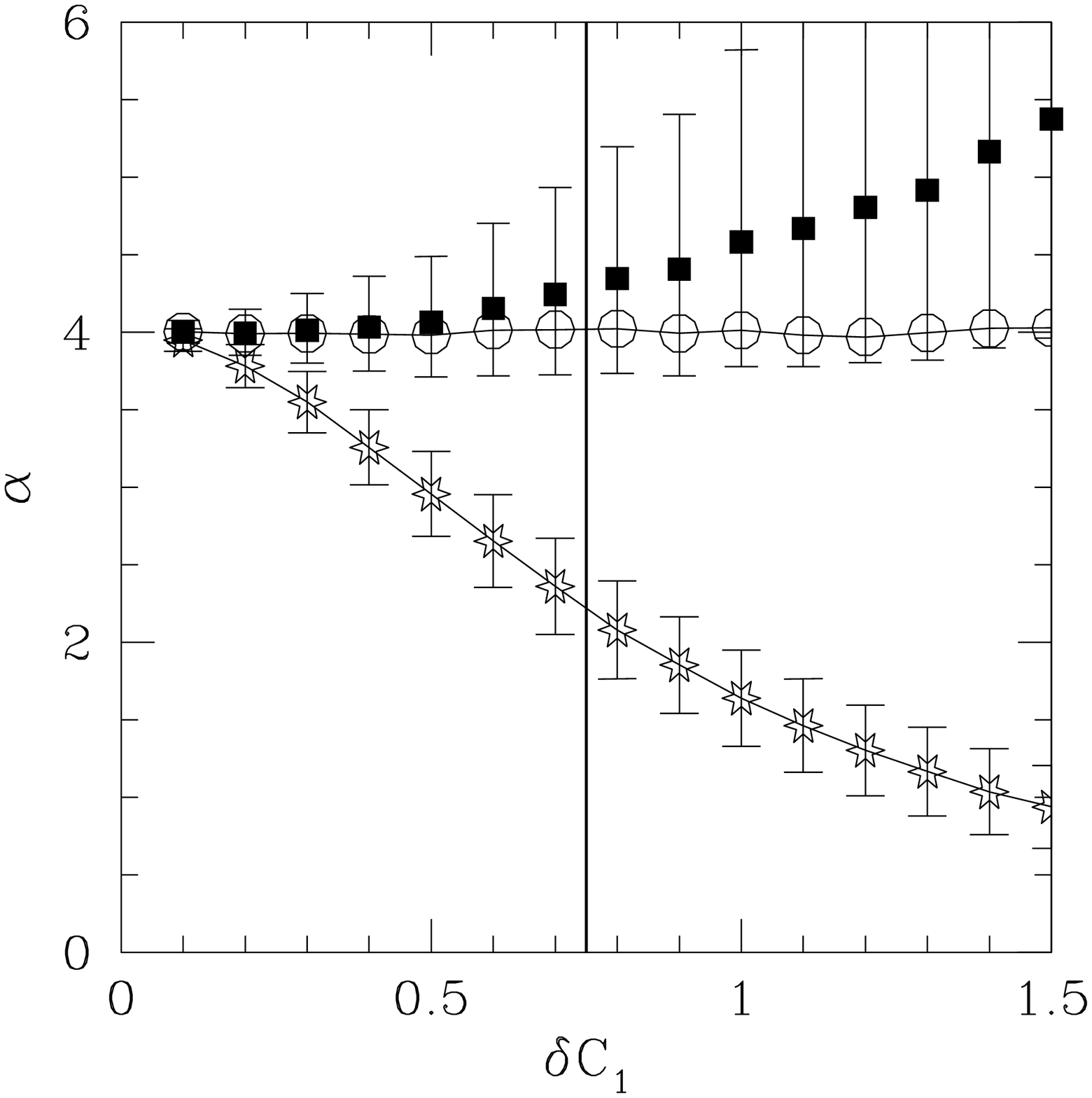}
\caption{(a) Manifestation of the statistical bias for 5000 ``groups''
that have been drawn from a $L_x \propto \sigma_v^4$ relation (solid
line), which
have a scatter of 0.8 in $\log L_x$. The recovered direct regression
line is the dashed one.
(b) Recovered values of the $L_x - \sigma_v$ slope
($\alpha$) for different amounts of scatter of the relation for the
direct regression ($\log L_x \propto \alpha \log \sigma_v$;
star symbols), for the inverse regression ($\log \sigma_v \propto
(1/\alpha) \log L_x$; open symbols) and for the inverse
regression once we have imposed an upper limit in $\sigma_v=1000$ km s$^{-1}$.}
\end{figure*}

As a further test of this bias, we have re-analysed the group data of Xue \& Wu
(2000). Their sample consists of 60 groups that have velocity
dispersion and X-ray luminosity data. Performing the direct regression
fit we recover their results (listed in their table 3 - OLS method),
ie. $\alpha\simeq 1$ and $C_1 \simeq 40$. However, if we perform an
inverse regression fit to their data we recover a very different
slope, ie., $\alpha \simeq 6.7$ and $C_1 \simeq 25.7$. In Fig.3 we plot
the $L_x - \sigma$ correlation for this sample and the fitted lines
for both regression methods. Using our Monte-Carlo procedure we have
seen that such a dichotomy between the direct and inverse regression
methods can be approximately accommodated if there is a scatter of
$\sim 1$ in $\log L_x$ and a cutoff at $\sigma\simeq 600$ km s$^{-1}$.
\begin{figure}[t]
\includegraphics[width=9cm]{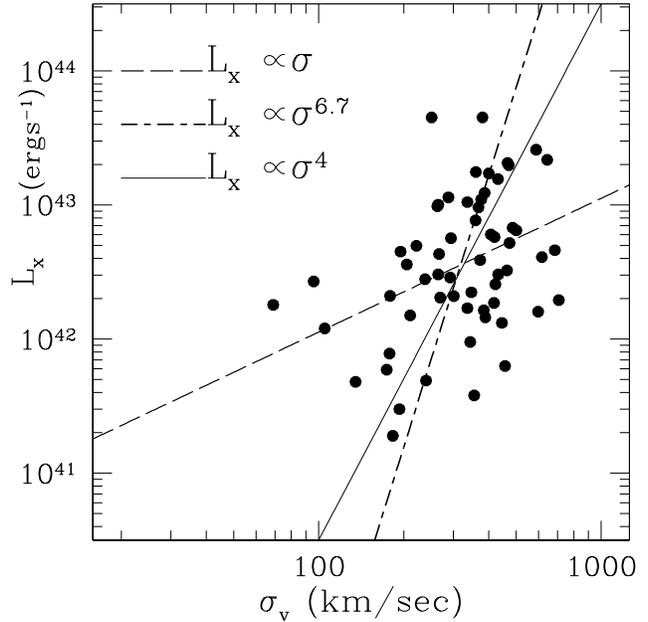}
\caption{Results of the Xue \& Wu (2000) sample of groups.}
\end{figure}

Guided by our Monte-Carlo analysis we performed an inverse regression
to our sample of $N\le 20$ groups, and we have indeed
recovered a slope 
$$\alpha \simeq 4.1 \pm 0.6 \;.$$
We therefore conclude that this sample of X-ray groups is absolutely
consistent with the relation found from clusters of galaxies.

\subsection{The scatter in the $L_x - \sigma$ relation}
Now we move to address the issue of the larger scatter of the $L_x -
\sigma_v$ relation of groups than of clusters. This is an important
issue because, as we have seen previously, the
larger scatter will induce an artificial
flattening of the $L_x - \sigma_v$ relation.

We will show below that the larger scatter 
is not only statistical, due to possible measurement uncertainties
especially in the low-$L_x$ and $\sigma_v$ regime, but
has a probably large intrinsic component which is due to an orientation effect.
To guide the reader through our arguments we need
first to define the axis ratio $b/a$ of the groups studied
\footnote{$a$ is the angular distance between the most widely separated 
galaxies in the group, and $b$ is the angular distances $b$ of the third 
galaxy of the group consisting of three galaxies from the line $a$ 
joining the most separated two galaxies, or is the sum of the angular 
distances $b_{1}$ and $b_{2}$ of the most distant galaxies on either side of 
the line $a$ joining the most separated galaxies (Rood 1979).}.
The ratio $b/a$ was determined by using the positions of member
galaxies which are mentioned in the corresponding references of Table
1 in Mulchaey et al. (2003). For those groups that Mulchaey et
al. (2003) used the NASA Extragalactic Database (NED), we selected 
the members of the corresponding groups by their 
redshifts taking into account the membership number, mentioned in Mulchaey et 
al. (2003). Note that that member galaxies are not drawn from a
well-defined and uniform magnitude limited sample. This can introduce
differences in the depth coverage within groups, but such an effect
would probably introduce a random error in the determination of the group
elongation.
In Table 1 we present $b/a$ values for those groups with number of 
members $N<20$. This is the sample we will investigate in detail. 

\begin{table*}[]
\hfill
\caption[]{The $b/a$ ratios and the number of galaxy members for 
groups with $2<N<20$.}
\tabcolsep 8pt
\begin{tabular}{|lrr|lrr|lrr|lrr|}
\hline
 Group & N & $b/a$ & Group & N & $b/a$ & Group & N & $b/a$ & Group & N
 & $b/a$ \\  
\hline
NGC 315  & 4 &0.53 & NGC 1407 & 8&0.36  & NGC 4065&7&0.38 & HCG 68 & 5& 0.53 \\
NGC 326  & 9 &0.56 & NGC 1587 & 4&0.03  & NGC 4104&8&0.66 & ARP 330 &8& 0.44 \\
NGC 524  & 8 &0.54 & NGC 2300 &13&0.58  & NGC 4125&3&0.44 & NGC 6338&11& 0.46 \\
HCG 12   & 5 &0.63 & HCG 37 &5 &0.39    & NGC 4261&8&0.26 & HCG 90 &16& 0.67 \\
NGC 720  & 4 &0.79 & HCG 48 & 3&0.08    & SHK 202&5&0.75 & UGC 12064&9& 0.79 \\
HCG 15   & 6 &0.49 & CGCG 154-041&4&0.13& NGC 4291&11&0.42 & HCG 92&4& 0.14 \\
HCG 16   & 9 &0.34 & NGC 3607 &7  &0.33 & NGC 4636&12&0.79 & IC 1459&5& 0.22 \\
UGC 1651 & 3 &0.02 & NGC 3647 & 6&0.24  & NGC 5044&9&0.63 & NGC 7619&7& 0.51 \\
NGC 1044 &13 &0.50 & NGC 3665 & 4 &0.19 & NGC 5171&15&0.26 & HCG 97 &14& 0.62 \\
IC 1880  & 7 &0.48 & HCG 57 & 7 &0.34   & HCG 67 &14 &0.73 & NGC 7777&4 & 0.63 \\
UGC 2775 & 5&0.39 & NGC 3923 & 5 &0.06  & NGC 5322& 8&0.36  &          &     \\
\hline
\end{tabular}
\end{table*}

We have noticed that there is a fairly significant correlation ($R=0.37$ and
${\cal P}=0.025$) between flattening and
velocity dispersion, with flatter systems having lower velocity
dispersions (see Fig. 4).
\begin{figure}[t]
\includegraphics[width=9cm]{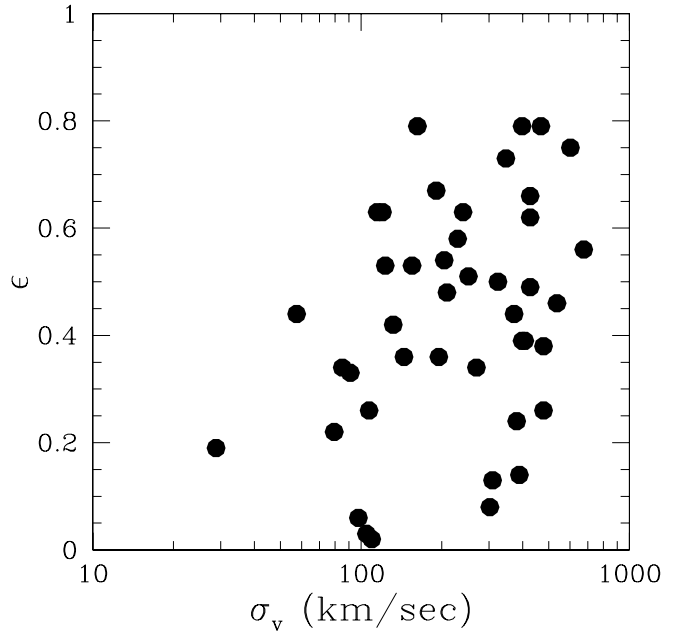}
\caption{Group axial ratio verus group $\log \sigma_v$.}
\end{figure}
Taking the 10 groups with the lowest
and the highest values of $\sigma_v$ ($\log\sigma_v \le 2.07$ and $\ge
2.6$, respectively) we find that the median and
68\% quantiles of their $b/a$ ratio is
$0.22^{+0.04}_{-0.16}$ and $0.49^{+0.07}_{-0.10}$, respectively. 

This difference in the group flattening is also accompanied by a 
significant difference in the number of group members, with mean
values of $\sim 5.2$ and 9.2 respectively. This 
could be expected either (a) due to the
number of galaxies - mass correlation (higher number of galaxies implies 
higher mass which then implies a higher velocity dispersion) or (b) due
to the fact that sparser groups are more likely, merely by chance, 
to show higher elongations than more dense systems.
The latter possibility has been ruled out after performing a large
set of Monte-Carlo simulations in which we have generated the same
number of groups and group members, as in the two observed subsamples,
by randomly placing group members in a sphere and then projecting them
to a plane. The resulting median axial ratio for the low and
high-$\sigma$ subsamples was found to be: $\sim 0.62 \pm 0.18$ and 
$\sim 0.7 \pm 0.15$, respectively, significantly larger than the
observed values.
However, neither the former possibility
can explain the significant difference in the flattening of
the two extreme subsamples of groups. Why should poorer groups be
flatter than richer ones ? A possible explanation could be the
different level of group virialization.
If groups accrete material anisotropically along one dimensional
structures, like filaments, then one may expect that flatter systems
are dynamically younger (thus they have lower values of $\sigma_v$ and
$L_x$) while when virialization takes place, after the groups have
accreted enough material, it will drive groups to more spherical
configurations, with higher velocity dispersions.

This straight-forward explanation cannot account however for
all the observables.
For example, selecting only the groups with $N\le 5$
members (in total 10 groups), which as expected from the above
discussion should be very flat (indeed their median $b/a$ is
$0.14^{+0.05}_{-0.06}$), we find that
although in this subsample there is no correlation of $\sigma_v$ with
the number of group members, there is a significant correlation
between their flattening and $\sigma_v$ (correlation coefficient R=0.8
and ${\cal P}_{random} = 0.006$), with low-$\sigma_v$ groups being
flatter. In the virialization paradigm
discussed above such a correlation cannot be explained. 
In order to explain such correlations we suggest that an {\em orientation
effect} is at work, i.e., if galaxy members move along their group 
major axis (for example, infalling in their common center of mass 
or rotating in elongated orbits around their gravitational center; 
cf. Tovmassian 2001, 2002), then flat groups oriented close to
orthogonally to the line of sight (small $b/a$) will have small values
of $\sigma_v$, while the opposite is true for flat groups seen edge-on.

According to this view the length of the apparent major axis, $a$, 
of the flattest groups (small $b/a$ ratio) should also depend on orientation. 
Groups oriented close to the line of sight will have 
on average small $a$ and high $\sigma_v$, and groups oriented close to the 
orthogonal to the line of sight will have large $a$ and small $\sigma_v$.
Tovmassian (2002) showed that such 
anti-correlation between the $a$ and $\sigma_v$ is 
observed in the case of HCGs groups. In order 
to investigate this possibility in our sample, we select the 11 groups 
with $b/a<0.35$. In Fig. 5 we plot their apparent value of $a$ versus
$\sigma_v$. The expected trend is indeed apparent, with $a$ increasing as
$\sigma_v$ decreases, and has a probability of occurring randomly of ${\cal P}=0.07$. 
Performing a bootstrap resampling technique in order to estimate the uncertainty of the 
above significance level, we find $\delta {\cal P}=0.02$. 

The existence of the orientation effect does not imply that the
virialization arguments, discussed previously, are incorrect. Most probably both are at work
since the orientation effect is not apparent in groups
with large number of members, exactly because these groups are richer
and most probably in a more advanced dynamical state.

\begin{figure}[t]
\includegraphics[width=9cm]{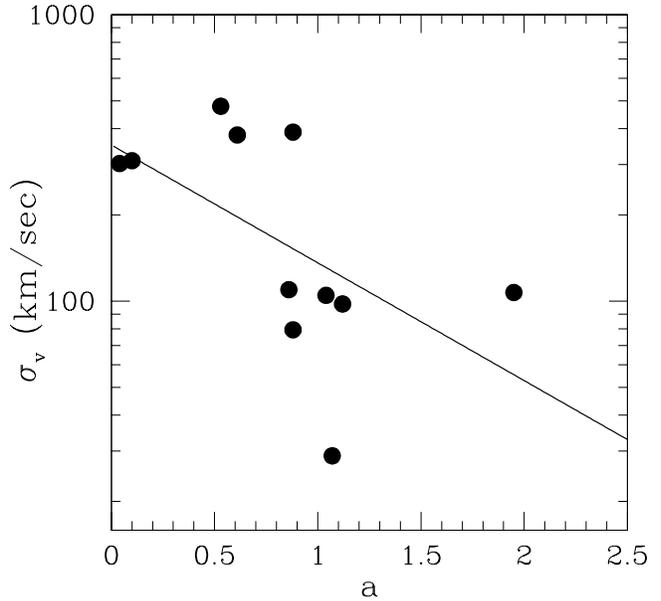}
\caption{The flat ($b/a<0.35$) group $a$ - log $\sigma_v$ correlation.}
\end{figure}

\section{Conclusions}
We have investigated the X-ray luminosity - velocity dispersion
relation in the new group Atlas of Mulchaey et al. (2003). A direct
regression shows that there is a strong 
correlation with a slope ($\alpha \simeq 1.7$) significantly shallower 
than that found in clusters of
galaxies. However, we attribute this to a statistical bias, resembling
the Malmquist bias, that enters in the direct regression approach and
which is due to a limit in $L_x$. We have quantified this using
Monte-Carlo simulations and once we use the more accurate inverse
regression method we obtain a slope of $\sim 4.1$, absolutely consistent
with the cluster relation. We have also investigated the apparently
larger scatter of the $L_x - \sigma$ relation of groups with respect
to clusters. We find that at least part of this scatter is intrinsic
in nature and
due to an orientation effect by which flat groups seen edge-on have
their $\sigma_v$ values underestimated. 
This can be understood by noting that groups of galaxies
have a roughly prolate spheroidal shape and thus if 
member galaxies move along their major axis, being accreted for example to
their common centre of mass, then the correlation between $\sigma_v$ 
and their major axis, could be due to the different group
orientations with respect to the line of sight.
Flat groups that are oriented roughly orthogonally to the line
of sight will have low values of $\sigma_v$ , while when oriented 
close to the line of sight or at intermediate angles, they will have 
higher values of $\sigma_v$.

\section*{Acknowledgments}
We thank the referee for suggestions that helped improve this paper.
MP acknowledges funding by the Mexican Government grant
No. CONACyT-2002-C01-39679.

\end{document}